# EVIDENCE FOR 1809 keV GAMMA-RAY EMISSION FROM $^{26}$Al DECAYS IN THE VELA REGION WITH INTEGRAL/SPI

**Stéphane Schanne[(1,*)], Patrick Sizun[(1)], David Maurin[(1)], Bertrand Cordier[(1)], Andreas von Kienlin[(2)], Clarisse Hamadache[(1)]**

*[(1)] CEA-Saclay, DAPNIA/Service d'Astrophysique, F-91191 Gif sur Yvette, France*
*[(2)] Max-Planck-Institut für extraterrestrische Physik, D-85740 Garching, Germany*
*[(*)] corresponding author: schanne{at}hep.saclay.cea.fr*

## ABSTRACT

The Vela region is a promising target for the detection of 1.8 MeV γ-rays emitted by the decays of radioactive $^{26}$Al isotopes produced in hydrostatic or explosive stellar nucleosynthesis processes. COMPTEL has claimed 1.8 MeV γ-ray detection from Vela at a 3σ level with a flux of 3.6×10$^{-5}$ ph/cm$^2$/s. In this paper, we present first results of our search for 1.8 MeV γ-rays from Vela with the spectrometer SPI aboard INTEGRAL. Using the data set acquired during 1.7 Ms at the end of 2005 in the frame of our AO-3 open-time observation, we determine a flux of $(6.5 \pm 1.9_{(stat)} \pm 2.4_{(syst)})\times10^{-5}$ ph/cm$^2$/s from $^{26}$Al decays in the Vela region.

## 1. VELA REGION NUCLEOSYNTHESIS

The Vela region is a key target for detecting gamma-rays emitted by the decays of radioactive isotopes produced in hydrostatic and explosive stellar nucleosynthesis processes, such as $^{26}$Al (gamma-ray energy: 1809 keV), $^{60}$Fe (1173 and 1332 keV) and the associated positrons, detectable through e$^+$-e$^-$ annihilation (511 keV). $^{26}$Al nuclei in particular, with a half-life of 0.74 Myr, can escape their production environment, before their decay to $^{26}$Mg associated with the emission of observable 1809 keV γ-rays.

The importance of the Vela region as a nucleosynthesis laboratory is underlined by the presence of potential nucleosynthesis sites (young Supernova Remnants (SNR), Wolf-Rayet (WR) stars) at distances as close as a few hundred parsec. Candidates for $^{26}$Al production in the Vela region are for instance: the Vela SNR (core collapse SN, age~10$^4$ yr, distance~250 pc) and the historical SNR Vela Jr (680 yr, ~250 pc), the WR star γ$^2$-Vel (~260 pc) and distant galactic OB associations (few kpc).

COMPTEL observations have revealed 1.8 MeV γ-ray emission from the Vela region [1] at a 3σ level, with a flux of (3.6±1.2)×10$^{-5}$ ph/cm$^2$/s, yet the source of the emission has so far not been identified.

In this paper, we present results of our search for $^{26}$Al gamma-ray emission in the Vela region with the spectrometer SPI aboard INTEGRAL [2], capable of resolving high-energy γ-ray lines (and particularly well in the 1.8 MeV domain) with good sensitivity thanks to its Ge-detectors.

## 2. NEW INTEGRAL VELA OBSERVATIONS

We use for our analysis the SPI data acquired in November-December 2005 in the frame of our accepted AO-3 open-time observation. Data were taken from INTEGRAL revolution 373 to 383, in 473 different satellite pointings, corresponding to 1.7 Ms of on-time. At variance with the previous deep Vela observations [3], during the AO-3 Vela data taking no strong solar flare polluted the data set (see Fig. 1). The SPI cryostat remained at a very stable temperature, without any cryocooler activity, hence the standard SPI energy calibration per revolution is sufficient for a line analysis with a few keV resolution at 1.8 MeV.

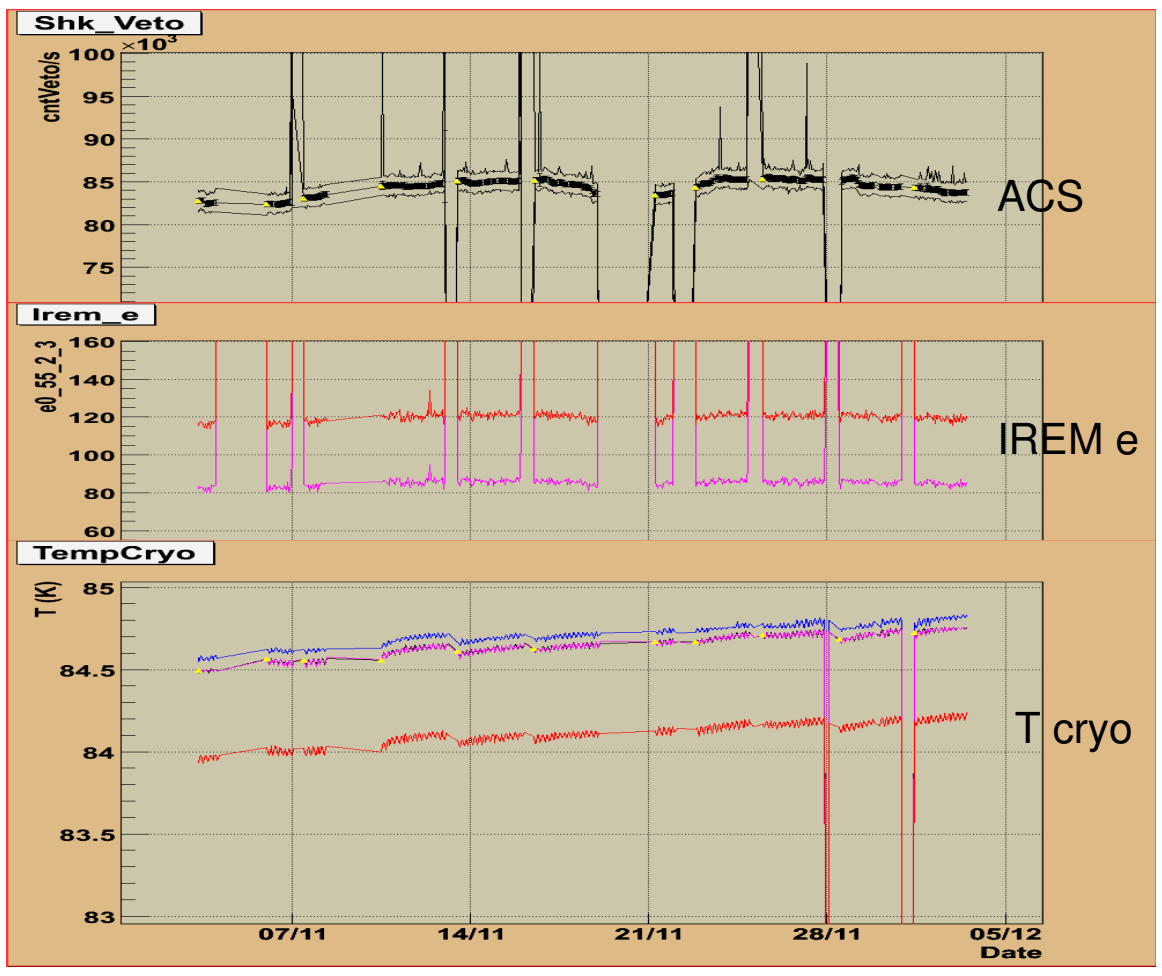

*Fig. 1. The count-rate profiles of the SPI anti-coincidence shield (ACS) and radiation monitor (IREM electrons) during data taking are smooth, and show only radiation belt entry/exits. Without cryomachine activity, the temperatures of the Ge cryostat increase only slowly.*

## 3. METHOD OF DATA ANALYSIS

We adopt a light-bucket approach, in which the counts of all active 17 SPI detectors are added together, ignoring the mask imaging properties. Background *tracers*, i.e. available dead-time-corrected count rates (such as saturating Ge rates, PSAC, IREM, and line rates), are used to build background models. Those are fit to an OFF dataset from which no $^{26}$Al signal is expected.. The background model is then applied to the ON database, using the tracers from the ON dataset, to predict and subtract the background present in the 1.8 MeV region.

For the OFF database we use all high-latitude (|B|>20°) SPI data since July 2004, from the moment on where only 17 working SPI detectors were left. For the ON database we use all public data since July 2004 and our private Vela AO-3 data. All data are cleaned in order to remove pointings of bad quality (containing radiation belt passages, or covering solar flares), following a procedure described in [4]. In particular cuts are applied to the cryogenic temperatures, PSAC, IREM proton (11-30 MeV) and electron (>.55 MeV) count-rates, before removal of spikes in the $^{12}$C, $^{16}$O, D lines and ACS count rates (using a boxcar algorithm). After cleaning, the OFF database contains 1697 pointings and the ON database 3560. For each pointing we extract a list of 259 relevant numbers (mostly dead-time corrected count rates, such as Ge saturating counts or ACS counts, or from selected regions in the Ge spectra, including e.g. background lines).

## 4. REFERENCE $^{26}$Al ANALYSIS

The spectrum of the SPI background in the 1.8 MeV region is shown in Fig. 2. The $^{26}$Al signal is searched in the single event spectra in a 10 keV wide bin centered at 1809 keV (which implies a narrow-line hypothesis, see Fig. 2). For the following reference $^{26}$Al analysis, we define also spectral regions (labeled *tracers 3,4,5,6*) covering the $^{205}$Bi and $^{28}$Al peaks (well correlating to the background in the signal region), and control bands (lo,hi) around the $^{26}$Al signal region. The background in the signal region is a combination of a continuum background (which can be deduced from the control bands) and a background peak from excited $^{26}$Na and $^{56}$Mn. We define also as *tracers 1,2,7* the Ge-saturating counter, the $^{69m}$Zn spectral band, and the PSAC.

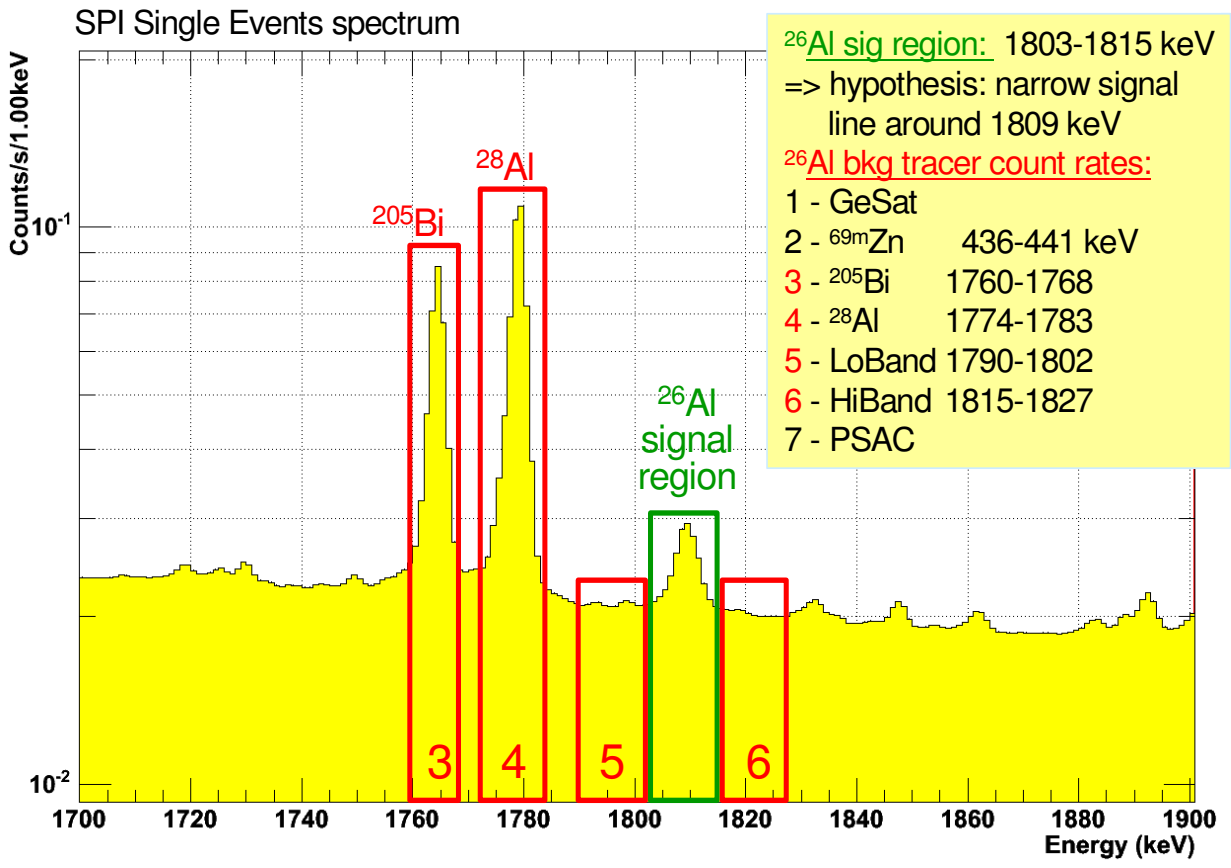

*Fig 2. SPI background in the $^{26}$Al signal region.*

All 7 *tracer* vectors (for the OFF and ON together) are then centered (in order to have zero mean values) and normalized, their covariance matrix diagonalized, their eigenvalues sorted from the most to the least significant. The new orthogonal set of base vectors forms the "principal components" (Fig. 3).

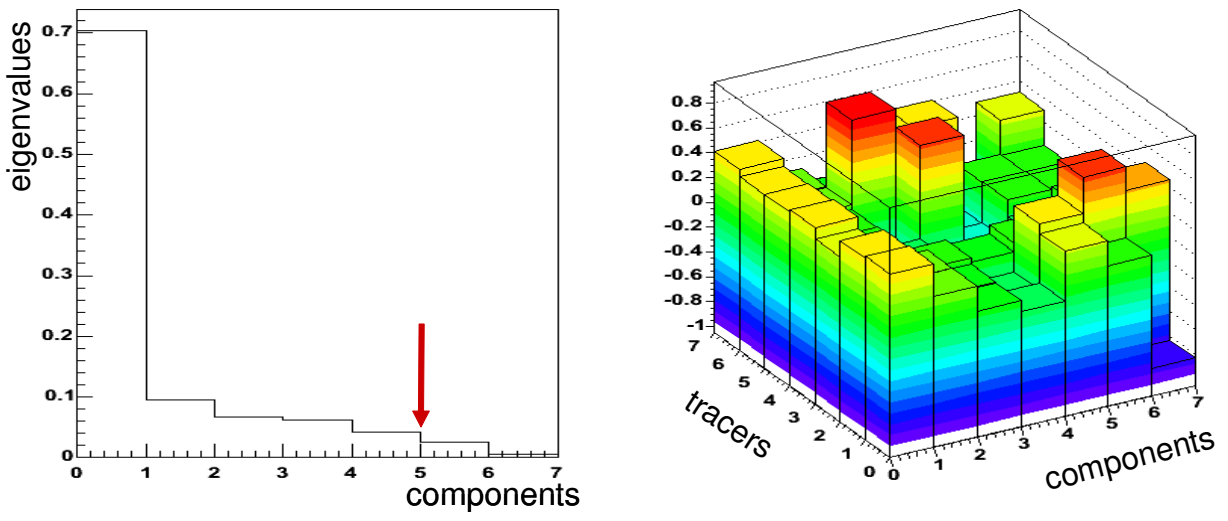

*Fig. 3. Eigenvalue of each principal component and matrix of the "principal components" as linear combination of the original (centered/normalized) tracers.*

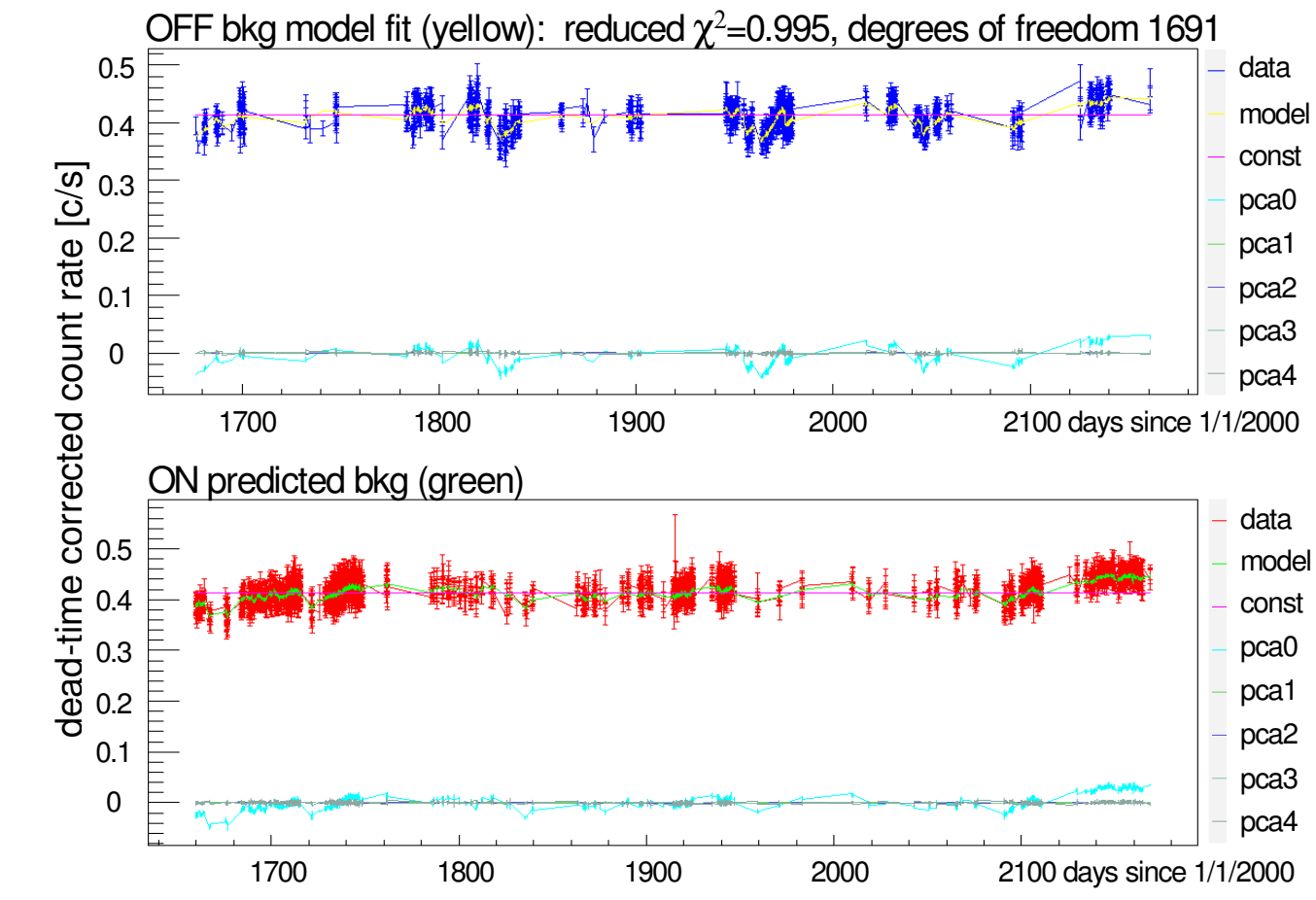

*Fig. 4. OFF/ON (top/bottom) count rates in the signal region vs time, superimposed with a linear background model fitted on the OFF data, and applied to the ON data. Each component used is shown with its weight $c_i$.*

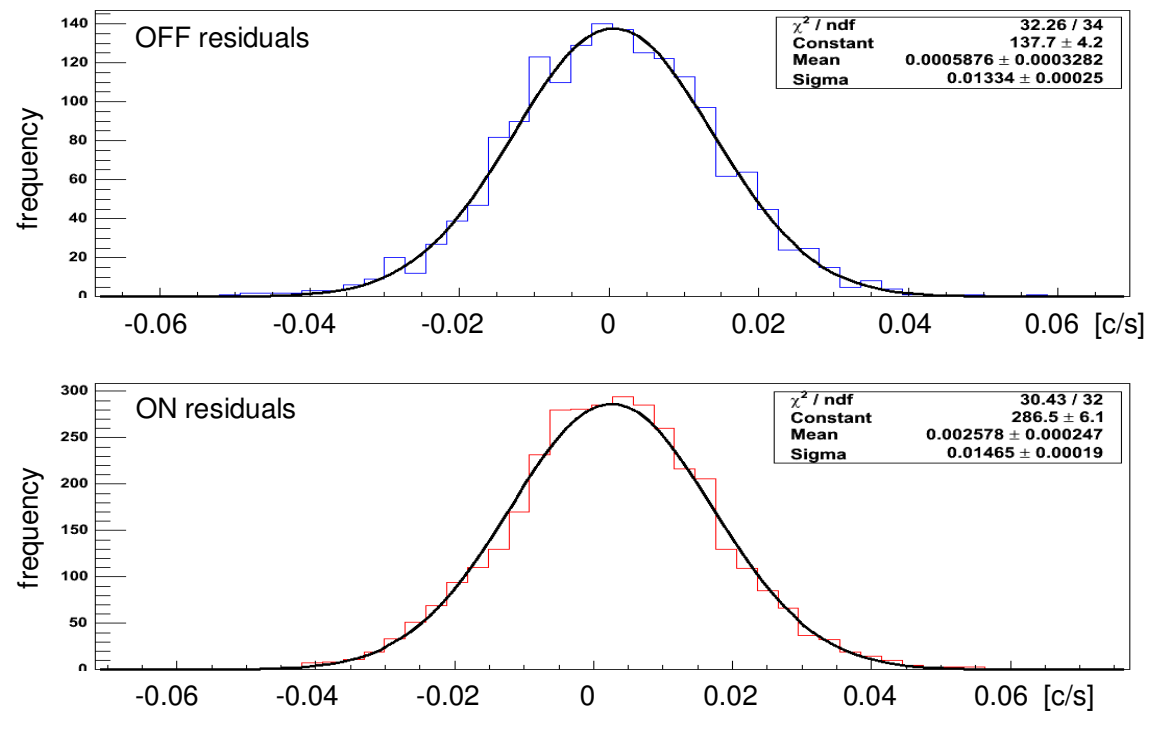

*Fig. 5. Histogram of the (dead-time corrected) residual count rate in the signal region after background subtraction. While the OFF histogram (top) is centered, the ON histogram (bottom) is shifted towards positive values, with a significantly non zero mean value, showing the detection of an overall Galactic plane $^{26}$Al emission.*

Using either all tracers or all components for the background model is equivalent. However by using only the first components it is possible to remove the noise accumulated in the components with the smallest eigenvalues. In the following we will apply the so-called Kaiser-cut, and use only the 5 first components for background modeling (note that in some cases even the 1st component alone gives already good results).

A linear background model is built using the OFF data pointings $p$ by fitting the coefficients $c_i$ to reproduce the (background) count rate $B_p$ in the signal region using the principal components $T_{i,p}$ as tracers:

$$B_p = c_0 + \sum_{i=1..Ncomponents} c_i \cdot T_{i,p}$$

This formula (with the $c_i$ found) is then used to predict, in the ON data, the background rate $B_p$ under the measured rate $R_p$ with the ON data tracers $T_{i,p}$ (shown in Fig. 4). The signal rate is then simply the residual $S_p=R_p-B_p$ (Fig. 5 and Fig. 6).

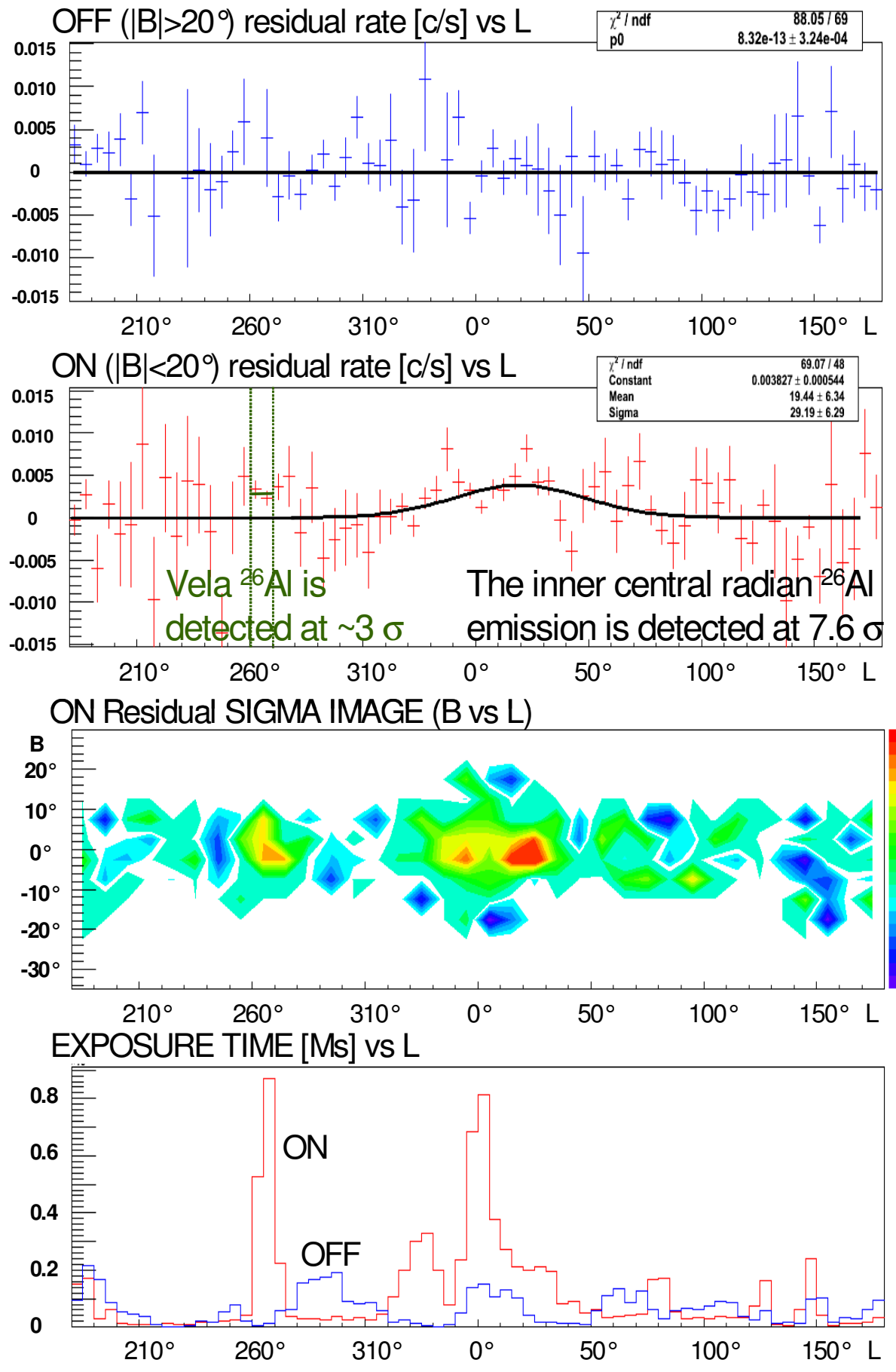

*Fig. 6. As a function of the Galactic longitude (L in °), are shown the OFF residual rate (fluctuating around 0), the ON residual rate (non zero indicates $^{26}$Al flux), the σ-significance map (latitude vs longitude) of the ON residuals, and the exposure (mostly Vela region and Galactic central radian).*

The background model fit on the OFF data is quite good, with a reduced $\chi^2$ of 0.995 for 1651 degrees of freedom. The results (i.e. the residual $^{26}$Al γ-ray fluxes) are presented in Fig. 6 as a function of the Galactic longitude. $^{26}$Al emission is significantly detected in the Galactic central radian (at a 7.6 σ level with the dataset used), and in the Vela region (with 4.4 σ).

The Galactic central radian 1809 keV γ-ray rate detected in this analysis is $3.8\pm0.5_{(stat)}$ counts/ks from SPI single events, for which the efficient area for a point source on axis is 32.8 cm$^2$ at 1.8 MeV. Assuming a uniform diffuse source with an extent of -30° to 30° in latitude, a reduction by 0.47 applies (SPI off axis efficiency integrated over this region). Hence we deduce a flux of $(2.4\pm0.35_{(stat)})\times10^{-4}$ ph/cm$^2$/s originating from $^{26}$Al in the Galactic central radian, in agreement with the COMPTEL value of $(2.8\pm0.15)\times10^{-4}$ ph/cm$^2$/s [5]. The Vela 1809 keV rate detected is $2.7\pm0.6_{(stat)}$ counts/ks from SPI single events, which translates to a flux of $(8\pm2_{(stat)})\times10^{-5}$ ph/cm$^2$/s, in the case of a point-like source (extent typically < 10°), only 2 σ above the COMPTEL value.

## 5. SYSTEMATIC $^{26}$Al STUDIES

In order to assess its systematic error, the previous analysis is repeated for all combinations of the following parameter variations:

(i) three different separations between ON/OFF data sets are chosen (|B|<10, 20, 30°);

(ii) three background model types are built (LNP=linear, with original tracers used for the background model; LP5=linear using the 5 first principal components as tracers, like presented in §4; BP5=Bayesian model, using the 5 first principal components – it replaces the linear model by the probabilistic neural network PNNV code given in [6], as described in [4], and illustrated in Fig. 7.);

(iii) five different combinations of tracers are used (the 7 tracers presented in §4, a list of 69 tracers from rate meters and single event lines, a list of 52 tracers reduced to strong lines, a list of 79 tracers with single and double event lines, and a list of 22 tracers containing only rate meters and no spectral tracers).

In this systematic test, all presented OFF background models have good fits: their $\chi^2$ values range from 0.9 for Bayesian models (an effect known as neural network "over-learning" from statistical data fluctuations) to 1.04 for the "worst" linear model. Therefore the $\chi^2$ value can not reasonably be used to discriminate the best models.

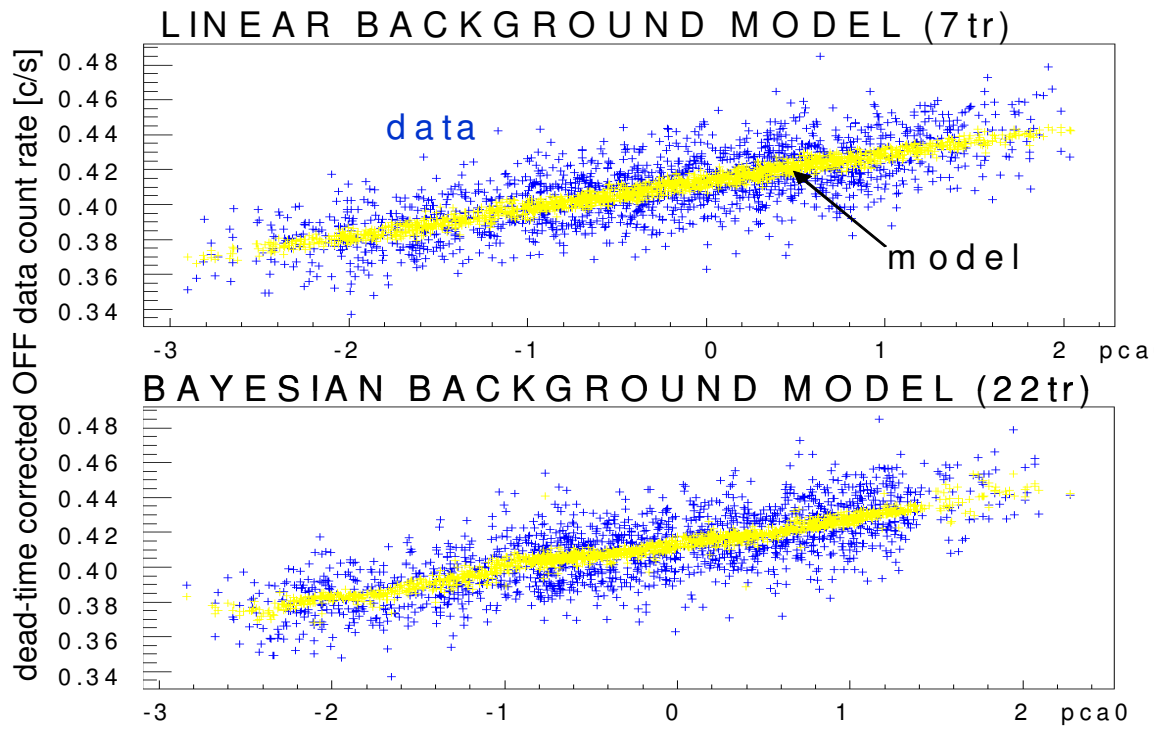

*Fig. 7. Normalized OFF data rate (blue) and model (yellow) vs first principal component rate for a linear and Bayesian model (which obviously allows for non linearity).*

The results of all combinations of systematic tests are shown in Fig. 8, presenting for each background model the 1809 keV flux obtained, which is computed using the measured rate and a SPI single event efficient area of 32.8 $cm^2$, i.e. the FOV central value is used. This is a correct assumption for a source with an extension<10° FWHM (FOV FWHM~24°).

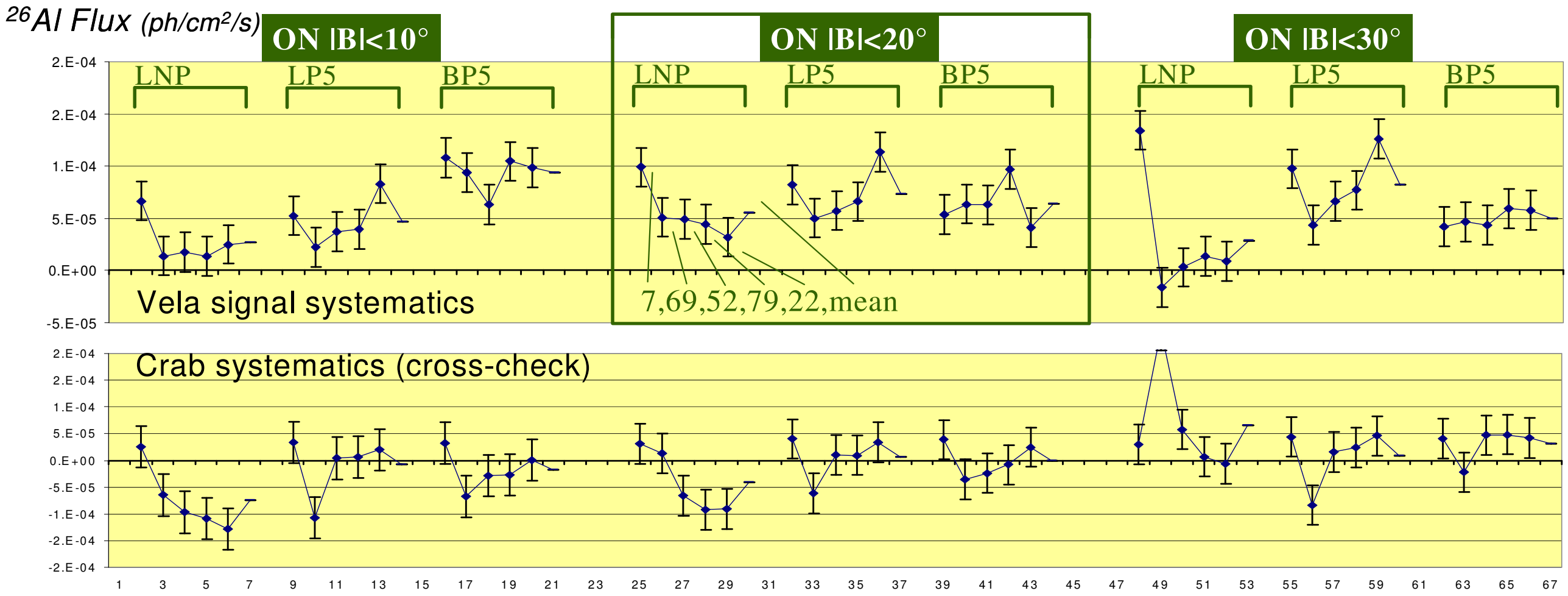

*Fig. 8. The 1809 keV γ-ray flux detected in Vela (top) and Crab (bottom) versus a sequential trial number, ordered in 3 groups (of ON regions |B|<10, 20, 30°) with 3 types (of background models LNP, LP5, BP5) using 5 different tracer lists (with 7, 69, 52, 79, 22 tracers). The error bars shown are 1σ-statistical. Each "-" marks the mean flux value for each group of 5 tracer lists.*

For an ON with |B|<30°, too few OFF data are left over to get a stable background model, hence the larger spread. For an ON with |B|<10°, there might be still signal in the OFF data (FOV HWHM~12°). Therefore we consider the ON with |B|<20° the best compromise. It gives indeed a lower systematic error. Notice also that in the case of the Crab, no signal is found and systematic errors are similar. The Bayesian background models show slightly lower systematic errors.

From this analysis, we obtain a detection of $^{26}$Al in the Vela region at 3.4 σ (statistical) with a flux, computed as the mean value for ON with |B|<20°, of $6.5 \pm 1.9_{(stat)}$ in units of $10^{-5}$ ph/$cm^2$/s, the systematic error in the same units being ±2.4 (standard deviation of all models) or +4.9/-3.2 (their min-max values).

## 6. CONCLUSIONS

From the analysis of 1.7 Ms of SPI data, corresponding to the first part of our Vela deep observation in the frame of INTEGRAL AO-3, taken during November-December 2005 (1.3 Ms more expected in 2006), we detect a 1809 keV γ-ray flux of $(6.5 \pm 1.9_{(stat)} \pm 2.4_{(syst)}) \times 10^{-5}$ ph/$cm^2$/s originating from $^{26}$Al decays in the Vela region. We considered all SPI pointings falling inside a L=10° and B=40° wide region centered at L=265° and B=0°, using SPI single events only, and assuming an efficient area of 32.8 $cm^2$ at 1.8 MeV (corresponding to the value on the SPI optical axis). This assumption is correct, only if the emission region is small in size compared to the SPI field of view. The systematic error is still large and the method has still to be carefully evaluated.

Our $^{26}$Al gamma-ray flux value from the Vela region is compatible with the one from COMPTEL. With our light-bucket approach we still can not disentangle the $^{26}$Al source, and an improved method, capable of using the mask imaging properties of SPI will have to be applied.

From the fact that we have an evidence for $^{26}$Al gamma-ray emission at the same level as COMPTEL, but in a 10 keV narrow energy bin, we can speculate that the 1.8 MeV line in Vela could be a narrow line. Further analysis is needed in order to measure the 1809 keV line width. If it turns out to be narrow, as the line observed from the Galactic central radian [7], this could then be a hint that the Vela region $^{26}$Al source might not be one of the recent and nearby supernova remnants present in the region, but rather be part of the overall Galactic $^{26}$Al emission. The localization of this source can be determined, with more statistics, using model fitting methods or SPIROS associated with a precise background model.